\documentclass[12pt,preprint]{aastex}

\shorttitle{Existence of the "radio thermally active" SNRs}
\shortauthors{Oni{\' c} et al.}

\begin{document}

\title{On the existence of "radio thermally active" Galactic supernova remnants}

\author{D. Oni{\' c}\altaffilmark{1},
  D. Uro{\v s}evi{\' c}\altaffilmark{1,2},
  B. Arbutina\altaffilmark{1}, and D. Leahy\altaffilmark{3}}

\altaffiltext{1}{Department of Astronomy, Faculty of Mathematics, University of Belgrade, Serbia}
\altaffiltext{2}{Isaac Newton Institute of Chile, Yugoslavia Branch}
\altaffiltext{3}{Department of Physics and Astronomy, The University of Calgary, Canada}

\begin{abstract}
In this paper, we investigate the possibility of significant production of thermal bremsstrahlung radiation at radio
continuum frequencies that could be linked to some Galactic supernova remnants (SNRs). The main targets for this
investigation are SNRs expanding in high density environments. There are several indicators of radio
thermal bremsstrahlung radiation from SNRs, such as a flattening at higher frequencies and thermal absorption at lower 
frequencies intrinsic to an SNR. In this work we discuss the radio continuum properties of 3 SNRs 
that are the best candidates for testing our hypothesis of significant thermal emission. In the case of SNRs IC443 and 
3C391, thermal absorption has been previously detected. For IC443, the contribution of thermal emission 
at 1 GHz, from our model fit is 3-57\%. It is similar to the estimate obtained from the thermal absorption properties 
(10-40\% at 1 GHz). In the case of the 3C391 the conclusions are not so clear. The results from our model fit 
(thermal emission contribution of 10-25\% at 1 GHz) and results obtained from the low frequency absorption 
(thermal contribution of 0.15-7\% at 1 GHz) do not overlap. For the SNR 3C396 we suggest that if previously detected 
thermal absorption could be intrinsic to the SNR then the thermal emission ($<$47\% at 1 GHz from our model fit) 
could be significant enough to shape the radio continuum spectrum at high frequencies. Polarization observations for these 
SNRs can constrain the strength of a thermal component. Reliable observations at low frequencies ($<100$ MHz) are needed 
as well as more data at high radio frequencies ($>1$ GHz), in order to make stronger conclusions about the existence of 
"radio thermally active" SNRs.
\end{abstract}


\section{Introduction}

The radio continuum emission from supernova remnants (SNRs) is believed to be mainly
produced by the non-thermal synchrotron mechanism. The radio continuum spectrum is
well fitted by a simple power law. On the other hand, the X-ray radiation from SNRs
is produced by thermal bremsstrahlung and line radiation as well as by non-thermal
synchrotron radiation (Reynolds 2008, and references therein). In this paper we investigate the
possibility of significant production of thermal bremsstrahlung radiation at radio frequencies
from SNRs that fullfill certain conditions, as discussed below.

{The approximate equation for the volume emissivity of thermal bremsstra\-hlung radiation $\varepsilon_{\nu}^{T}$ 
for an optically thin ionized gas cloud at radio frequencies has the form:}
\begin{equation}
\varepsilon_{\nu}^{T}=6.8\times10^{-38}\ g_{ff}(\nu, T)\ n^{2}\ T^{-0.5}\
[\mathrm{erg\ cm^{-3}\ s^{-1}\ Hz^{-1}}],
\end{equation}
where the number densities of electrons and ions are the same and given by $n$
in $\mathrm{cm}^{-3}$, the temperature of the emitting region $T$ is in $\mathrm{K}$ and the thermally averaged Gaunt factor $g_{ff}(\nu, T)$ 
at radio frequencies is given by Gayet (1970) as:
\begin{equation}
g_{ff}(\nu, T)\approx\left\{
\begin{array}{ll}
0.55\ln(4.96\times10^{-2}\nu^{-1})+0.82\ln{T}, & 10^{2}\ \mathrm{K}<T<9\times10^{5}\ \mathrm{K} \\
0.55\ln(46.80\nu^{-1})+0.55\ln{T}, & T\gtrsim9\times10^{5}\ \mathrm{K}
\end{array}
\right.
\end{equation} where frequency $\nu$ is in $\mathrm{GHz}$.

From equation (1) we see that if the number density increases, the thermal volume emissivity increases (with the
square of density). On the other hand, as temperature increases, the thermal volume emissivity decreases
(approximately as $T^{-0.5}$). {Figure 1 shows the thermal bremsstrahlung volume emissivity for a range of radio
frequencies as a function of number density ($1-10^{3}\ \mathrm{cm}^{-3}$) and temperature ($10^{4}-10^{5}$ K). In the 
case of the post Sedov-Taylor phase of SNR evolution, especially for those SNRs expanding in high
density environment, the thermal radio volume emissivity will increase with time as temperature decreases and density
increases.}

\begin{figure}[!t]\centering
  \includegraphics[width=\columnwidth]{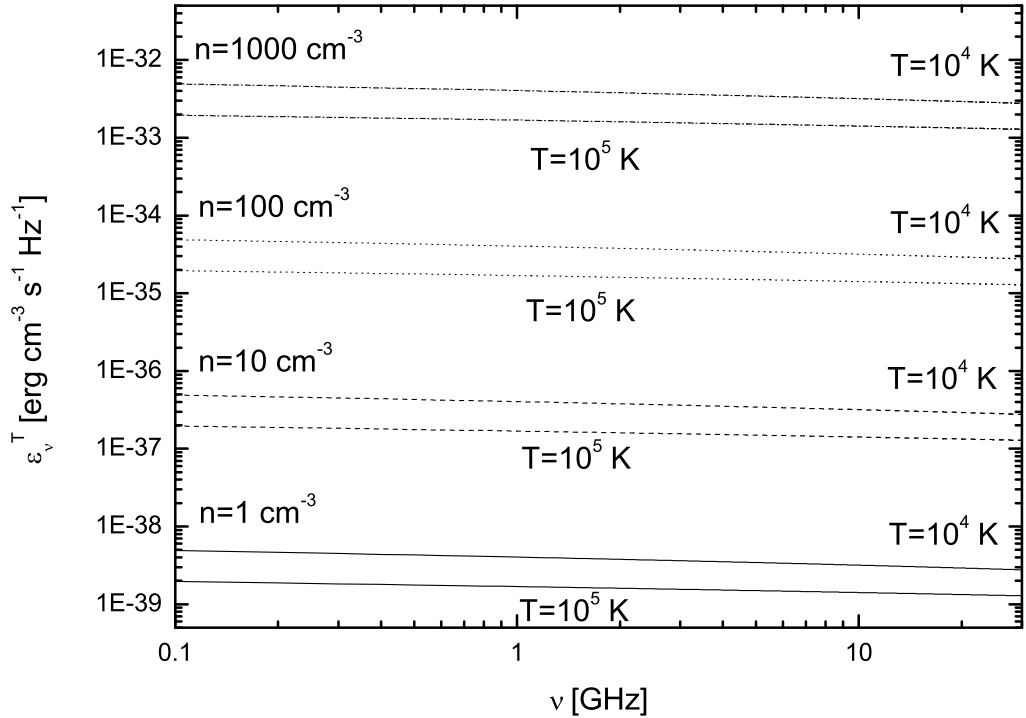}
  \caption{Thermal bremsstrahlung volume emissivity as a function of radio frequency, for
different number densities and temperatures.}
  \label{fig:one}
\end{figure}

The SNR shock wave ionizes, heats and compresses gas in the SNR environment. An SNR embedded in a high density
environment will become evolutionary old sooner than one which expands in the rarefied ISM. The main hypothesis that 
we discuss in this paper is: if an SNR expands into a high density interstellar medium (ISM), e.g.\@ a molecular cloud 
environment, significant thermal bremsstrahlung emission in the radio continuum from the SNR should be produced during its 
evolution.

In this paper, we analyze the possibility of significant thermal bremsstrahlung
emission at radio frequencies from SNRs (Section 2). Furthermore, we propose methods for
estimating the contribution of the thermal bremsstrahlung component in the total volume
emissivity at 1 GHz and discuss the results for some possible "radio thermally
active" Galactic SNRs (Sections 3 and 4). Due to their nature, we exclude from our analysis
filled-center or Crab-like SNRs.

\section{Theoretical overview}

\subsection{Radio continuum spectra}

The spectra of SNRs in radio continuum are usually represented by a power law (representing pure synchrotron radiation 
from the SNR shell). For frequency in units of GHz, the flux density can be represented by the following expression: \begin{equation}
S_{\nu}=S_{1\ \!\!\mathrm{GHz}}\cdot\nu^{-\alpha}\ [\mathrm{Jy}],
\end{equation} where: $S_{1\ \!\!\mathrm{GHz}}$ is the spatially-integrated flux density at 1 GHz, and $\alpha$ is the radio
spectral index.

{Observed} radio spectral indices vary from around 0.3 to 0.8 for Galactic SNRs (see Green
2009). Test-particle diffusive shock acceleration (DSA) theory predicts that for strong shocks, 
the radio spectral index is approximately 0.5. In the case of shocks with Mach
number less than around {ten}, steeper spectra are expected; on the other hand, few SNRs
would be expected to have such weak shocks {(Reynolds 2011)}. Steeper radio spectra may also be explained
by nonlinear effects. An important prediction of the modified shock explanation,
where the reaction effects of cosmic ray (CR) particle pressure is taking into account, is
that the spectrum should flatten at higher energies so that a "concave up" spectrum
is formed (see Reynolds 2008 and references therein). In the case of young SNRs,
where the acceleration process is efficient, there is some evidence for spectral
curvature, although better measurements with smaller errors are needed
(Allen, Houck \& Sturner 2008, Reynolds \& Ellison 1992). {Bell, Schure \& Reville (2011) noted that the total 
steepening may be a cumulative effect of non-linear and oblique-shock steepening in the case 
of young SNRs.} On the other hand, observations over a very broad range of
radio frequencies reveal a curvature in the spectra of some evolutionary older Galactic
SNRs (Uro{\v s}evi{\' c} \& Pannuti 2005, Tian \& Leahy 2005, Leahy \& Tian 2006, Uro{\v s}evi{\' c}, Pannuti \& Leahy 2007, 
Oni\'c \& Uro{\v s}evi{\'c} 2008).

{When the spectral index is changing with frequency there is a possibility that two different mechanisms are producing 
the radio emission. The presence of thermal bremsstrahlung radiation would change the shape of the radio spectrum to be
"concave up", especially in the case of SNRs evolving in a dense environment (evolutionary old SNRs).}

If we take the approximative formula for the
Gaunt factor at radio frequencies from Cooray \& Furlanetto (2004) and references therein: $g_{ff}(\nu,
T)\approx11.96T^{0.15}\nu^{-0.1}$ we can write for the sum of thermal
bremsstrahlung ($\varepsilon^{T}_{\nu}$) and non-thermal synchrotron
($\varepsilon^{NT}_{\nu}$) emissivities:
\begin{equation}
\varepsilon_{\nu}=\varepsilon^{NT}_{\nu}+\varepsilon^{T}_{\nu}=\varepsilon_{1\
\!\!\mathrm{GHz}}^{\mathrm{NT}}\nu^{-\alpha}+\varepsilon_{1\ \!\!\mathrm{GHz}}^{\mathrm{T}}\nu^{-0.1}.
\end{equation}
where frequency $\nu$ is in GHz. For optically thin emission we can write a similar formula
for the total flux density: \begin{equation}
S_{\nu}=S^{NT}_{\nu}+S^{T}_{\nu}=S_{1\ \!\!\mathrm{GHz}}^{\mathrm{NT}}\nu^{-\alpha}(1+x\nu^{\alpha-0.1}).
\end{equation} where $x=S_{1\ \!\!\mathrm{GHz}}^{\mathrm{T}}/S_{1\ \!\!\mathrm{GHz}}^{\mathrm{NT}}$ is the ratio of the thermal and non-thermal
flux densities at 1 GHz.

{As radio frequency increases, both thermal and non-thermal emissions are decreasing. The non-thermal 
spectrum is steeper than the thermal bremss\-tra\-hlung spectrum. If significant thermal radiation is emitted then the slope of 
the radio spectrum will change (concave up spectrum) at higher radio frequencies ($>$1GHz).} {This is clearly seen in 
Figure 2, which illustrates the case for a synchrotron spectral index of 0.5 and different values of $x$.} Another 
phenomenon, seen in Figure 2, is that the value of the power law index obtained by fitting the total spectrum will have a 
lower value than the intrinsic synchrotron spectral index. The relative change in the radio spectral index is 
$y=\alpha_{\mathrm{total}}/\alpha_{\mathrm{synch.}}$, where $\alpha_{\mathrm{total}}$ represents the radio spectral index 
estimated from the power law fit and $\alpha_{\mathrm{synch.}}$ represents the synchrotron spectral index.

\begin{figure}[!t]\centering
  \includegraphics[width=\columnwidth]{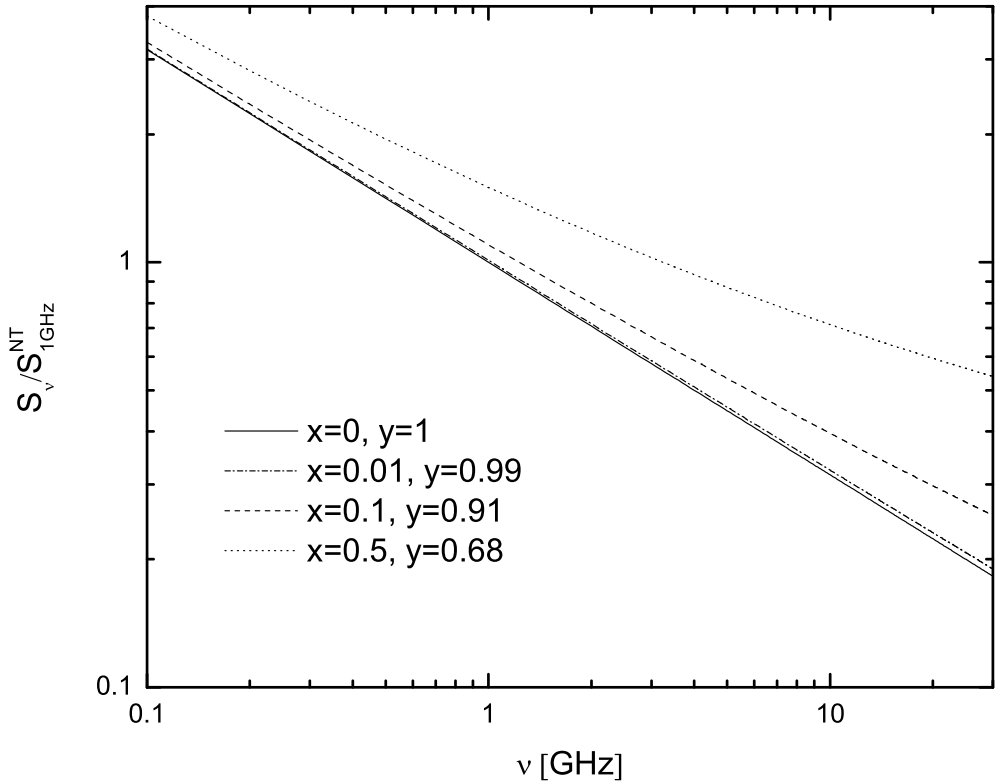}
  \caption{Curvature in an SNR radio spectrum due to existence of thermal bremsstrahlung emission 
in the case of the synchrotron spectral index of 0.5 and different values of $x$.}
  \label{fig:two}
\end{figure}

{It must be noted that there is a possibility that different areas of the SNR have different non-thermal spectra 
(for example, because of different evolutionary stages). If the spectral indices are not changing over the SNR and spectral 
flattening exists at higher frequencies, that could be indication of significant thermal emission from the SNR 
(assuming that there is no contamination by overlapping thermal sources). On the other hand, 
radio spectral index variations, detected in some SNRs and associated with parts of the SNR where density is higher 
than average (inhomogeneous medium with high density cloudlets, high density gradients due to vicinity of the 
molecular cloud) could be explained by intrinsic SNR thermal emission. However, the non-association of high 
linear polarization percentages with those regions is essential for validation of this claim.}

There is also a significant number of Galactic SNRs with $\alpha<0.5$. Contamination with flat spectrum 
thermal emission may be responsible for the lower spectral index values. It is also possible, in the case
of such a flatter spectra, that second-order Fermi (or stochastic) acceleration plays a major role (Reynolds 2008 and
references therein). Schlickeiser \& F\"{u}rst (1989) concluded that the observed dispersion in spectral index values
below $\alpha=0.5$ is attributed to a distribution of low plasma $\beta$ values ($\beta\simeq0.05$) in different remnants
($\beta=\frac{8\pi P}{B^{2}}\propto\left(\frac{V_{\mathrm{sound}}}{V_{A}}\right)^{2}$, where $P$ is the gas pressure,
$B$ magnetic induction, $V_{\mathrm{sound}}$ adiabatic sound speed and $V_{A}$ Alfv$\mathrm{\acute{e}}$n speed).
Ostrowski (1999) pointed out that shock waves with Alfv$\mathrm{\acute{e}}$n speed non-negligible in comparison to
the shock velocity are responsible for generation of the flat particle distribution. The corresponding analysis was applied 
to the SNRs W44 and IC443. The flatter spectra may also be due to compression ratios greater than four at radiative shocks 
(Bell et al.\@ 2011). It must be noted that there are no predictions of curved - "concave up" radio spectra in the 
cases of the mentioned explanations of the flat radio spectra seen in some SNRs.

One of the characteristics of the so called mixed-morphology (thermal composite)
SNRs\footnote{The mixed-morphology SNRs are with bright interiors in X-rays, and bright rims in radio.}
(Rho \& Petre 1998, Vink 2012) is that they usually have $\alpha<0.5$ based on a pure synchrotron fit (Green 2009). On the
other hand, they are known to expand in a high density environment (many of them interacting with molecular clouds) and
are mainly evolutionary old. Tilley, Balsara \& Howk (2006) emphasized that SNRs expanding into denser ISM might spend a
significant fraction of their observable lives as mixed-morphology remnants. {That marks mixed-morphology remnants as
the best candidates for the investigation of the possible production of significant thermal bremsstrahlung radiation at
radio frequencies. On the other hand, Uchiyama et al.\@ (2010) proposed that the radio emission may be additionally
enhanced by the presence of secondary electrons/positrons, i.e.\@ the products left over from the decay of charged pions,
created due to cosmic ray nuclei colliding with the background plasma. The presence of secondary electrons/positrons may
also explain the flat spectral radio indices of some mixed-morphology SNRs (Uchiyama et al.\@ 2010). {This model, on the 
other hand, do not produce a "concave up" radio spectrum.}
}

\subsection{Thermal bremsstrahlung radiation in the radio continuum}

Here, we investigate the hypothesis that in the case of SNRs in the late and post
Sedov-Taylor phases it is likely that the reason for a curved (concave up) radio
continuum spectrum is a significant presence of thermal bremsstrahlung radiation. {We discuss situation in which the 
ensemble of thermal electrons, linked to an SNR, could significantly radiate at radio frequencies.}

First, we consider the case where DSA is the only process of acceleration. The fraction
of protons and electrons in a gas that are pre-accelerated to an energy at which
they are injected into the acceleration mechanism is around $10^{-3}$
({Bell 1978a,b}, Berezhko \& V$\mathrm{\ddot{o}}$lk 2004). This leaves the possibility of the existence of a great pool of thermal electrons that
can radiate, so we can talk about two populations of electrons: thermal and
relativistic. {The number of relativistic electrons is always negligible compared to the thermal electrons, but thermal 
electrons can have temperatures of a $10^{6}$ K.} {Vink et al.~(2010) and Drury et al.~(2009) pointed out that the post-shock (downstream)
gas temperature could be significantly reduced by acceleration. For young SNRs and
Mach numbers $M>75$, the downstream temperature is at least 80 times the upstream
temperature (Vink et al.~2010). One of the consequences of efficient cosmic ray acceleration is that the highest energy 
cosmic rays may escape far upstream (i.e.~into the unshocked medium) forming a cosmic-ray shock precursor. Non-linear cosmic 
ray acceleration leads to lower plasma temperatures, in the most extreme cases perhaps even quenching thermal X-ray emission 
(Vink 2012, and references therein). Directly upstream of the shock the cosmic rays provide a non-negligible pressure, 
which pre-compresses and pre-heats the plasma (Vink 2012). The most important phases for particle acceleration are the 
transition into Sedov-Taylor phase, and early Sedov-Taylor phase (Reynolds 2008). The radio synchrotron luminosity increases 
with time in the free expansion phase, achieves its peak value at the very beginning of the Sedov phase, and then again 
decreases with time (Berezhko \& V$\mathrm{\ddot{o}}$lk 2004).} We suggest that, even when DSA is not too weak, there can be 
significant thermal bremsstrahlung radiation at radio continuum frequencies from SNRs expanding in a high 
density environment with lower temperatures.

It is known that in the case of radiative J shocks, a photo-ionized precursor is formed.
Half of the photons produced in the cooling zone downstream to the shock, pass
upstream, where they are absorbed by the upstream gas. {The ambient material becomes ionized so if
its density is high, preserving the optically thin medium, there is the possibility of significant thermal bremsstrahlung
radiation at radio continuum frequencies.} As Brogan et al.~(2005) concluded in the case of free-free
absorption observed toward SNR 3C391, ionized gas can be produced when the SNR blast
wave encounters a nearby molecular cloud with sufficient speed to dissociate and ionize the gas.

Pynzar' \& Shishiv (2007, and references therein) analyzed thermal emission and
absorption in and near directions towards SNRs and used it to estimate the distribution
of ionized gas surrounding SNRs of type II supernovae. They concluded that the kinetic
energy of a type II supernova envelope is transformed into radiation at an early
stage in the evolution of the object, leading to the formation of an \mbox{H\,{\sc ii}} region in the vicinity of the
SNR or young pulsar. This could mean that the thermal radio emission is inherent to SNRs.

In the case of older SNRs, it is not yet known whether the mechanism for producing relativistic electrons is local acceleration at a shock front or compression of the
existing population of Galactic background relativistic electrons (Leahy 2006 and references therein). A variable magnetic 
field strength is expected in older SNRs. The compression ratio in radiative shocks can be very large, resulting in a 
stronger and more strongly variable magnetic field. Analyzing the spectral index variation in the case of SNR HB21, 
Leahy (2006) pointed out that the variable magnetic field causes the observed frequency range to
correspond to variable energy range in the electron spectrum, so that either positive or negative
curvature in the electron spectrum results in variable observed spectral index in a fixed
radio frequency range. Leahy \& Roger (1998) analyzed radio spectral index variations for the
Cygnus Loop and pointed out that a radio spectrum with concave up (positive) curvature
can be produced by adding two different emission spectra along the same line of
sight or within the same emitting volume (possibly due to two different electron populations).

\subsection{The low-frequency turnovers}

Detected concave down (negative) spectral curvature at low frequencies is believed
to be due to absorption by a thermal plasma, although the possibility of synchrotron self-absorption is not ruled out.
The low-frequency turnovers have been usually attributed to free-free absorption in ionized thermal gas along the line of
sight to the SNR, although it could be also directly linked to some SNRs (e.g.~3C391, IC443, see Brogan et al.~2005 and
{Castelletti et al.~2011}). Thermal absorption, at low frequencies, inherent to an SNR, could be a sign of possible
thermal emission at high frequencies.

Since thermal absorption could be present at lower frequencies, we estimate the
thermal cutoff frequency ($\nu_{\mathrm{T}}$) dependence on electron temperature ($T_{e}\ \mathrm{[K]}$) and emission
measure ($EM\ \mathrm{[cm^{-6}\ pc]}$) and present it in Figure 3. {We can
write (Altenhoff et al.~1960):} \begin{equation}
\nu_{\mathrm{T}}\approx0.3045T_{e}^{-0.643}EM^{0.476}\ \mathrm{[GHz]},
\end{equation} where: \begin{equation} EM=\int_{0}^{s}n_{e}^{2}ds.
\end{equation}

{For e.~g.~IC443 we calculated $\nu_{\mathrm{T}}\approx$(30-50) MHz for an assumed $T_{e}$=(8000-12000) K and
$EM=(2.8-5)\times10^{3}\ \mathrm{cm^{-6}\ pc}$ (eastern rim, {Castelletti et al.~2011)} and 3C391
$\nu_{\mathrm{T}}\approx$(40-150) MHz for an assumed $T_{e}$=(1000-3000) K
and $EM=(0.6-2.5)\times10^{3}\ \mathrm{cm^{-6}\ pc}$ (Brogan et al.~2005). In the case of 3C391, 
Anantharamaiah (1985) suggested the upper limits for $EM=2500\ \mathrm{cm^{-6}\ pc}$ and $T_{e}=8000$ K.}

\begin{figure}[!t]\centering
  \includegraphics[width=\columnwidth]{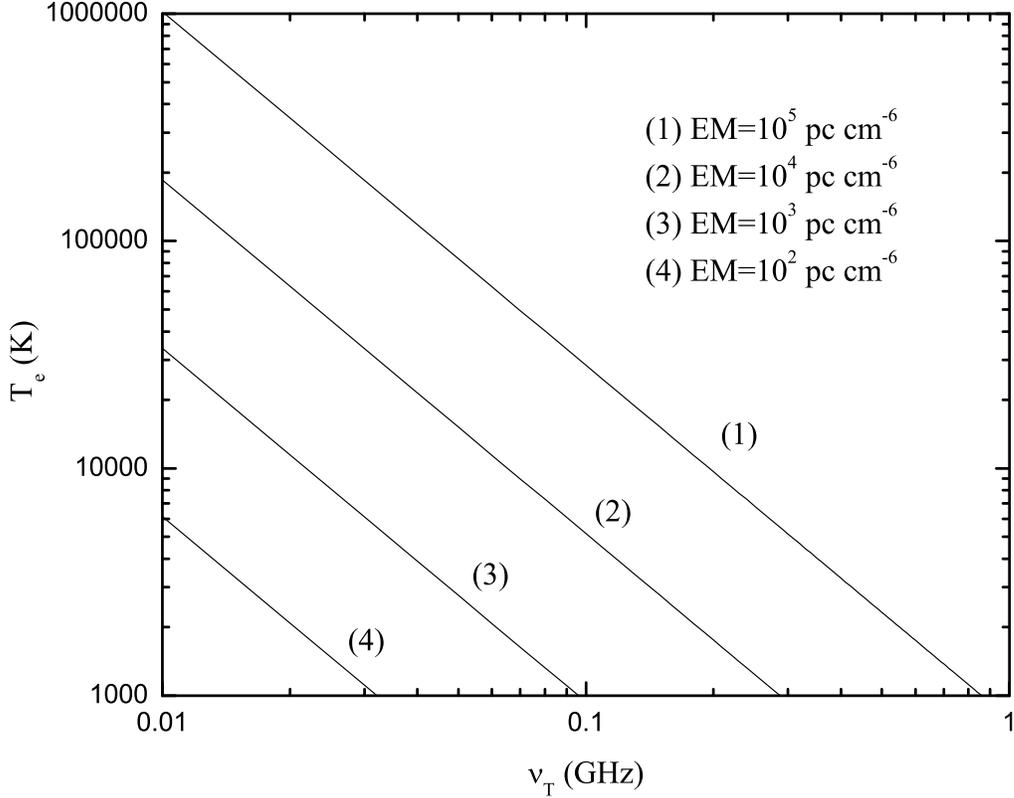}
  \caption{The thermal cutoff frequency $\nu_{\mathrm{T}}$ dependence on
electron temperature $T_{e}$ and
emission measure $EM$.}
  \label{fig:three}
\end{figure}

{If the low-frequency turnover due to thermal absorption is inherent to an SNR, we can calculate the flux density at the turnover
frequency $\nu_{T}$ from the Rayleigh-Jeans formula: \begin{equation}
                                           S_{\nu_{T}}=\frac{2kT_{e}\Omega_{S}\nu_{T}^{2}}{c^{2}}
                                          \end{equation}
and then calculate the flux density at any higher frequency from: \begin{equation}
                                                S_{\nu}=S_{\nu_{T}}\left(\frac{\nu}{\nu_{T}}\right)^{-0.1}
                                               \end{equation}
where $k$ is the Boltzmann constant, $c$ is the speed of light, $\Omega_{S}$ is the source solid angle,
and $T_{e}$ is the electron temperature. If there is an estimate of $T_{e}$, we can compare the flux density at 1GHz 
obtained from the low frequency absorption for a given SNR with the directly measured integrated flux density.
In Figure 4 the thermal flux density versus frequency is plotted, for different values of $\nu_{\mathrm{T}}$, $T_{e}$ and
$\Omega_{S}$ based on the equations (8) and (9). Solid lines are for $\Omega_{S}=10^{-6}$ sr while dotted lines are for
$\Omega_{S}=10^{-4}$ sr. Low-frequency free-free thermal absorption may provide an important complementary tracer
of fast ionizing SNR/molecular cloud shocks (for details see Brogan et al.~2005) as well as of thermal bremsstrahlung 
emission.

\begin{figure}[!t]\centering
  \includegraphics[width=\columnwidth]{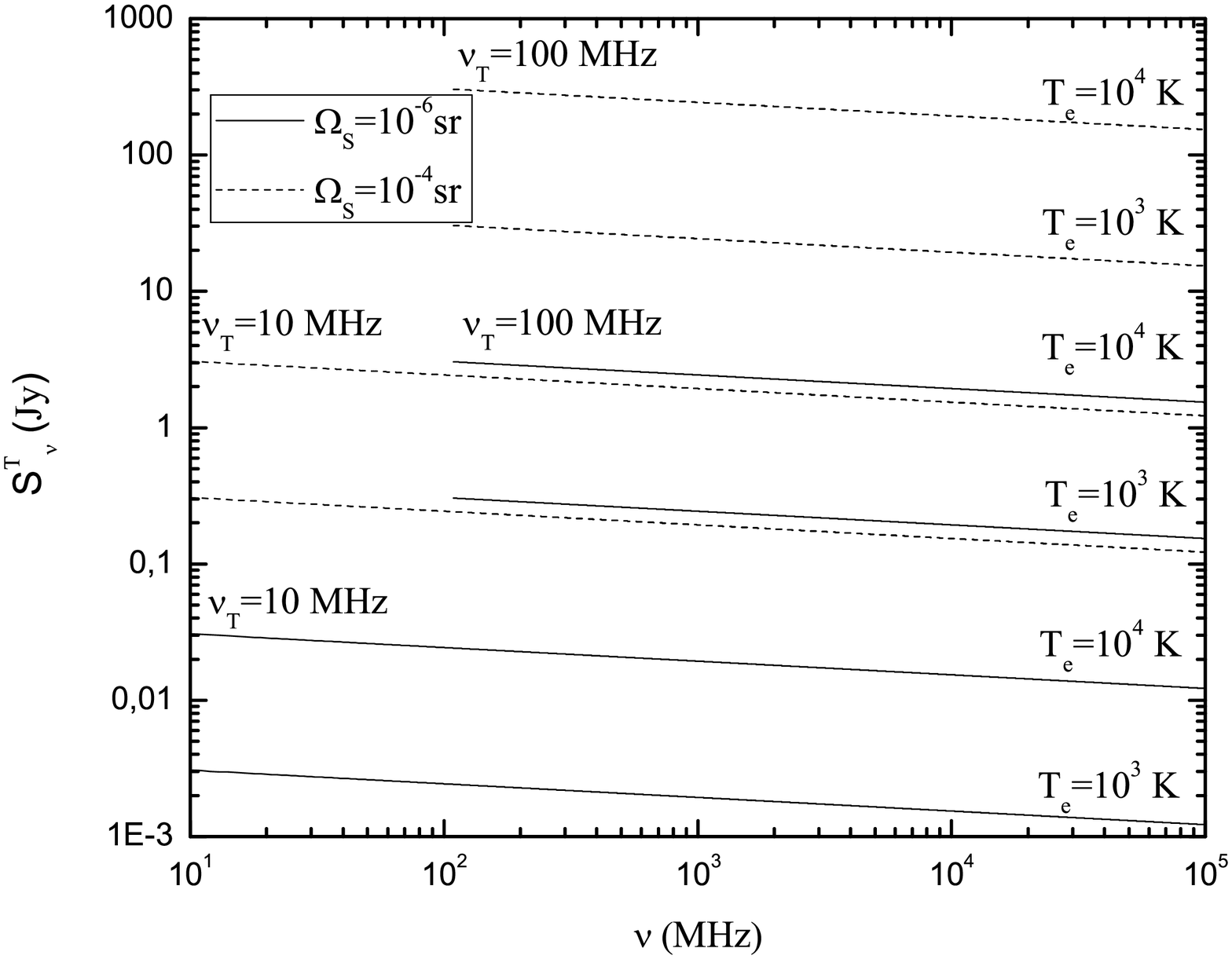}
  \caption{The thermal flux density versus frequency for different values of turnover frequency $\nu_{\mathrm{T}}$,
electron temperature $T_{e}$ and source solid angle $\Omega_{S}$ based on the equations (8) and (9).}
  \label{fig:four}
\end{figure}
}

\subsection{SNR dynamical evolution and thermal bremsstrahlung emission}

In the Sedov-Taylor phase, the downstream temperature $T$ (where the radiation is
generated) and the shock radius can be expressed by
(Yamauchi et al.~1999): \begin{equation}
T=2.07\times10^{11}\left(\frac{E_{51}}{n_{a}}\right)^{0.4}t^{-1.2}\ [\mathrm{K}],
\end{equation}
\begin{equation}
R_{S}=0.315\left(\frac{E_{51}}{n_{a}}\right)^{0.2}t^{0.4}\ [\mathrm{pc}],
\end{equation} where the age is expressed in years and the initial explosion energy is $E$: $E_{51}=\frac{E\
[\mathrm{erg}]}{10^{51}}$. Petruk (2005) noted that the transition time $t_{\mathrm{tr}}$ is defined as the end of
the energy conserving stage and the beginning of the radiative stage (so the Sedov-Taylor approximation is valid
before $t_{\mathrm{tr}}$). The shell formation time $t_{\mathrm{sf}}$  marks the time when the so-called pressure-driven snowplow
(PDS) phase defined by McKee \& Ostriker (1977) starts, in which interior hot gas
pushes the cold dense shell (the "deceleration parameter", defined as $m=d\log
R_{S}/d\log t$, is 2/7). The structure of the flow is re-structurized and a thin shell is formed during the so
called transition sub-phase given by the time interval ($t_{\mathrm{tr}}, t_{\mathrm{sf}}$). Petruk (2005) gave equations for
$t_{\mathrm{tr}}$ and $t_{\mathrm{sf}}$: \begin{equation}
t_{\mathrm{tr}}=2.84\times10^{4}E_{51}^{4/17}n_{a}^{-9/17}\ \mathrm{[yr]}
\end{equation}
\begin{equation}
t_{\mathrm{sf}}=5.20\times10^{4}E_{51}^{4/17}n_{a}^{-9/17}\ \mathrm{[yr]}
\end{equation}

{The transition time $t_{\mathrm{tr}}$, the shell formation time $t_{\mathrm{sf}}$, the shock radius $R_{\mathrm{tr}}$, the
temperature $T_{\mathrm{tr}}$ and the shock velocity $V_{\mathrm{tr}}$ for SNRs which expand in environments with
different ambient number densities $n_{a}$, with $E_{51}=1$, are given in Table 1.} It is clearly seen that SNRs which
expand in higher density environments are smaller and younger when they leave the energy conserving phase. SNRs
which expand in a high density environment evolve (and leave the Sedov-Tayler phase) faster then those in a low density 
environment. The approximate analytical solutions are not adequate for the post Sedov-Tayler phases, so that numerical 
simulations are needed.

\begin{table}
 \begin{center}
  \caption{Transition and shell formation time as well as radius, temperature and shock velocity at the transition time for SNRs expanding in the environments
with different ambient densities with $E_{51}=1$.} \label{tab:fitst}
 \begin{tabular}{cccccc}
    \tableline\tableline
    $n_{a}\ [\mathrm{cm^{-3}}]$ & $t_{\mathrm{tr}}\ [\mathrm{yr}]$ & $t_{\mathrm{sf}}\ [\mathrm{yr}]$ & $R_{\mathrm{tr}}\
[\mathrm{pc}]$ & $T_{\mathrm{tr}}\ [\mathrm{K}]$ & $V_{\mathrm{tr}}\ [\mathrm{km/s}]$ \\
    \tableline
  0.01 & $3.25\times 10^{5}$ & $5.95\times 10^{5}$ & 126.56 & $3.21\times 10^{5}$ & 152 \\
  0.1 & $9.61\times 10^{4}$ & $1.76\times 10^{5}$ & 49.03 & $5.53\times 10^{5}$ & 199 \\
  1 & $2.84\times 10^{4}$ & $5.20\times 10^{4}$ & 19.00 & $9.5\times 10^{5}$ & 261 \\
  10 & $8.39\times 10^{3}$ & $1.54\times 10^{4}$ & 7.36 & $1.63\times 10^{6}$ & 342 \\
  100 & $2.48\times 10^{3}$ & $4.54\times 10^{3}$ & 2.85 & $2.81\times 10^{6}$ & 449 \\
  1000 & $7.33\times 10^{2}$ & $1.34\times 10^{3}$ & 1.11 & $4.83\times 10^{6}$ & 584 \\
    \tableline
  \end{tabular}
\end{center}
\end{table}

{Weakening of the SNR shock in the post-Sedov-Taylor phase lowers the temperature of the shell as well as of the 
thermal X-ray emitting region (therefore reducing the extent of that region). The Sedov-Taylor phase ends 
when the shock is slow enough (usually $\sim200\ \mathrm{km\ s^{-1}}$) that significant radiative cooling can take place and the
adiabatic approximation breaks down. Significant radio thermal bremsstrahlung emission could arise from the cooled thermal X-ray electrons. 
On the other hand, local regions of the blast wave may become radiative sooner where the ambient density is much higher than average, 
though the bulk evolution is still adiabatic (Reynolds 2008). Different parts of an SNR may be in different phases. SNRs could have both radiative and 
non-radiative shocks. This could also give a rise to the significant radio thermal bremsstrahlung
emission from the parts of an SNR in the radiative phase. For example, the Cygnus Loop, earlier in its evolution, must have 
expanded rapidly within the tenuous medium inside the wind-blown bubble. Vink (2012) pointed out that, currently, the largest 
part of the shock of Cygnus Loop, is interacting with a dense shell swept up by the progenitor's wind. Another example is 
RCW 86 which also shows a mixture of radiative and non-radiative shocks (Vink 2012).

As already mentioned, mixed-morphology SNRs are characterized by centrally-peaked thermal X-ray emission with 
shell-like radio morphology. {Their interiors are more or less homogeneous in temperature and, because of pressure 
equilibrium, also reasonably homogeneous in density (Vink 2012)}. The Sedov-Taylor model is not 
applicable for mixed-morphology SNRs and their observational properties differ from those expected for an evolved SNR in a 
homogeneous ISM (Yamaguchi, Ozawa \& Ohnishi 2012). More complex models are needed to explain the mechanism responsible for 
the observed characteristics of mixed-morphology SNRs. Radiative recombination continuum (RRC) was recently discovered in 
their X-ray spectra (Yamaguchi, Ozawa \& Ohnishi 2012, and references therein). The presence of strong RRC is evidence that 
the plasma is recombining. Most of the mixed-morphology SNRs appear to be in the radiative phases of their evolution, 
with shock velocities less than 200 $\mathrm{km\ s^{-1}}$ (Vink 2012), which results in cool, X-ray dim, but optical/UV 
bright regions immediately behind the shock front. It is likely that a significant pool of cooled thermal electrons exist. 
{Vink (2012) noted that the high density in the interior of mixed-morphology SNRs is a direct consequence of the high 
ISM density and thermal conduction, which results in a more uniform interior density with medium hot temperatures, rather 
than very high temperatures for the case of very low interior densities.}

\subsection{The thermal radio luminosity and possibility of ambient density estimation from radio continuum emission}

Luminosity and flux density are related by:
\begin{equation}
L_{\nu}=4\pi d^{2}S_{\nu}.
\end{equation}
where $d$ is the distance to an SNR. {The thermal luminosity, on the other hand, can be expressed as:}
\begin{equation}
L_{\nu}^{T}\approx\varepsilon_{\nu}^{T}V_{\mathrm{shell}}=2\times10^{18}g_{ff
}(\nu, T)n^{2}T^{-0.5}\frac{4\pi}{3}R_{S}^{3}f,
\end{equation} where $L_{\nu}^{T}$ is in $[\mathrm{erg\ s^{-1}Hz^{-1}}],\ n
[\mathrm{cm^{-3}}],\ T[\mathrm{K}],\ \mathrm{shock\ radius}\ R_{S}
[\mathrm{pc}]$ and $f$ is the volume filling factor.

In the Sedov-Taylor phase, the compression ratio is nearly equal to 4 so we can
approximately write $n\approx4n_{a}$, where
$n_{a}$ is the average ambient number density {($\rho_{\mathrm{ISM}}=n_{a}\mu{m_{H}}$, where
$\rho_{\mathrm{ISM}}$ is the ambient density,} $\mu$ is the
mean molecular weight and $m_{H}=N_{A}^{-1}$, where $N_{A}$ is Avogadro's
number). We can also roughly assume for the volume filling factor of the shell: $f\approx0.25$. For the 
Sedov-Taylor phase, the thin shell approximation is only moderately accurate (Bandiera \& Petruk 2004).

{For a standard value of $E_{51}=1$ using equations (2) and (15), in the case of
the Sedov-Taylor phase (using equations (10) and (11)), we can write:}
\begin{equation}
L_{1\ \!\!\mathrm{GHz}}^{T}\approx\left\{
\begin{array}{ll}
1.66\times10^{15}fn^{2.5}_{a}R^{4.5}_{S}(16.99-0.82\ln{n_{a}}-2.48\ln{R_{S}}), &
T<9\times10^{5}\ \mathrm{K} \\
1.66\times10^{15}fn^{2.5}_{a}R^{4.5}_{S}(14.54-0.55\ln{n_{a}}-1.65\ln{R_{S}}), &
T\gtrsim9\times10^{5}\ \mathrm{K}
\end{array}
\right.
\end{equation}

From Truelove \& McKee (1999) we can estimate the shock radius at the beginning of
the energy conserving phase (approximately for the case of uniform ambient medium and
ejecta):
\begin{equation}
R_{ST}\approx2.23\left(\frac{M_{ej}}{M_{\odot}}\right)^{1/3}n^{-1/3}_{a}\
\mathrm{[pc]}
\end{equation}

An $L^{T}_{1\ \!\!\mathrm{GHz}}-R_{S}$ plot can also give us an
insight into the type of the SNR phase if we could calculate the $L^{T}_{1\ \!\!\mathrm{GHz}}$ from the observations
and our model fit (see Figure 5).

\begin{figure}[!t]\centering
  \includegraphics[width=\columnwidth]{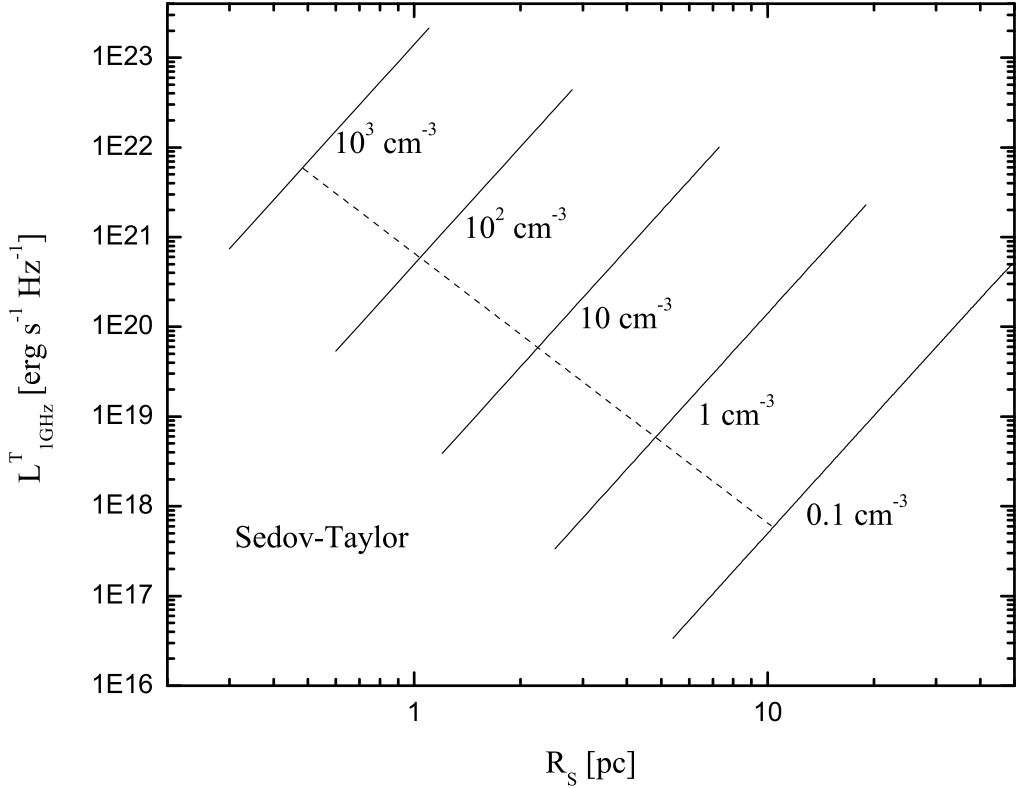}
  \caption{$L^{\mathrm{T}}_{\mathrm{1GHz}}$ vs.~$R_{S}$ in the case of a Sedov-Taylor SNR for different values of ambient
density. The endpoints of the thermal luminosity lines for pure Sedov-Taylor phase represent the moment when an SNR reach $R_{\mathrm{tr}}$. The dashed line
connects points that represent the beginning of the Sedov-Taylor phase if $M_{ej}=10M_{\odot}$ in contrast to
the $M_{ej}=1.4M_{\odot}$ assumed for the shown starting points.}
  \label{fig:five}
\end{figure}

{It is worth comparing Figure 5 with a similar graph for pure synchrotron emission (see Chapter 5 and Figure 4 in
Berezhko \& V$\mathrm{\ddot{o}}$lk 2004). In the classical Sedov-Taylor phase, the luminosity of thermal radiation is
{$\approx4$} orders of magnitudes less than of synchrotron radiation so it does not significantly contribute
to the overall radio continuum spectrum. We used $f=0.25$ so that different volume
filling factor values could change the results by factors of a few. The endpoints of the thermal luminosity lines for
Sedov-Taylor phase (see Figure 5) represent the moment when an SNR reaches $R_{\mathrm{tr}}$ (see Table 1). The dashed line
in Figure 5 connects points that represent the beginning of the Sedov-Taylor phase (equation 17) if $M_{ej}=10M_{\odot}$. 
This contrasts with $M_{ej}=1.4M_{\odot}$ assumed for the shown starting points (Truelove \& McKee 1999). During the 
energy conserving phase and at the beginning of the radiative phases the contribution of thermal radio emission is negligible. 
We can see from Table 1 that the temperature at the end of Sedov-Taylor phase is around $10^{5}-10^{6}$ K and it is nearly 
independent of explosion energy and ambient density. To decrease this temperature significantly either the explosion energy 
or the ambient density must be unrealistically low. If we can detect significant thermal X-ray emission the temperature has 
to be of order of $10^6$ K or at least a few $10^{5}$ K and then the radio bremsstrahlung emissivity would be to low.

As we already pointed out, the evolution of an SNR could be more complicated than that given by the classical
Sedov-Taylor model. In the case when an SNR expands in inhomogeneous medium, local regions of the blast wave may become
radiative sooner where the ambient density is much higher than average, though the bulk evolution is still governed by
Sedov-Taylor phase. The high density regions of an SNR populated by cooled thermal X-ray electrons could account for
significant radio thermal luminosity ($L_{\nu}^{T}\propto n^{2}$) that could be comparable with synchrotron emission. In
Sedov-Taylor phase $L_{\nu}^{NT}$ is approximately independent of the density (Berezhko \& V$\mathrm{\ddot{o}}$lk 2004).
At a given SNR diameter, all SNRs in Sedov-Taylor phase should have roughly the same synchrotron spectral luminosity, 
with some spread due to explosion energy. On the other hand, SNRs in denser media reach the Sedov time (corresponding to 
their peak luminosity) when they still have relatively small diameters. Therefore, their peak spectral luminosity is 
brighter than that of SNRs in lower density media, and they continue to be more luminous through much of the Sedov-Taylor 
phase (Chomiuk \& Wilcots 2009, and references therein). However, for radiative shocks, the shock compression factors 
are large, giving rise to strongly compressed magnetic fields, and enhanced cosmic-ray electron densities, so the radio 
synchrotron emission could be also strongly enhanced (Vink 2012, and references therein). We emphasize the importance 
of the analysis of evolution (and radiation) of SNRs which include more complex models of SNR dynamics.
}

\section{The model}

In order to distinguish the contribution of thermal and non-thermal components in the
total emission, the SNR radio integrated spectrum can be fitted by a sum of
these two components (error-weighted least squares). For frequencies in $\mathrm{GHz}$, the relation
for the integrated flux density can be written as follows: \begin{equation}
S_{\nu}=S_{1\ \!\!\mathrm{GHz}}^{\mathrm{NT}}\ \left(\nu^{-\alpha}+\frac{S_{1\
\!\!\mathrm{GHz}}^{\mathrm{T}}}{S_{1\
\!\!\mathrm{GHz}}^{\mathrm{NT}}}\ \nu^{-0.1}\right)\ [\mathrm{Jy}],
\end{equation} where: $S_{1\ \!\!\mathrm{GHz}}^{\mathrm{T}}$ and $S_{1\ \!\!\mathrm{GHz}}^{\mathrm{NT}}$ are flux densities 
corresponding to thermal and non-thermal components, respectively. We used weighted (instrumental errors) non-linear least 
square for fitting.

The whole SNR is assumed to be optically thin at radio frequencies and the thermal emission has spectral index equal to 
0.1. It is assumed that the synchrotron radiation is not absorbed or scattered by the thermal gas. {Also 
the radio spectral index is assumed to be constant in the SNR shell. The most important distinction between the presence of 
thermal emission or simply a particle acceleration model that produces a flat spectrum is the fact that the spectral index 
changes with frequency, a "concave up" spectrum is formed, because the low frequencies are dominated by steep spectrum 
synchrotron emission and the high frequencies by flat spectrum thermal emission. SNRs that do produce significant thermal 
emission are in a later evolutionary stage. As we have already mentioned, these SNRs can have different parts in different phases 
of evolution, with possibility of producing synchrotron emission with different spectral indices in different areas. Quite 
naturally, the flatter spectrum would dominate the high frequency part and the steeper spectrum the low frequency regime. 
This is one of the major problems of our model. To avoid that problem it has to be shown that the spectral index is not 
changing very much over the SNR. This is very difficult, in practice, with observations from various archives, observed 
at a wide frequency range, at various resolutions. On the other hand, as we already noted, radio spectral index variations, 
associated with parts of an SNR where the density is higher than average and for which linear polarization percentages are 
negligible, could be explained by intrinsic SNR thermal emission.}

This model is valid in the approximation of constant density and temperature in
the emitting shell. The model itself also assumes a simple sum of non-thermal and
thermal components. We note that this model, in general, does not distinguish between intrinsic
and foreground/background thermal emission.

In the case of SNR expansion in the vicinity of an \mbox{H\,{\sc ii}} region, the thermal emission
from an SNR and adjacent \mbox{H\,{\sc ii}} region can not be separated within our model. This is
a serious problem especially since we are interested in the analysis of the integrated SNR
spectrum. In the case when there is an overlapping \mbox{H\,{\sc ii}} region, it is very difficult to separate the 
thermal contribution from the \mbox{H\,{\sc ii}} region from the SNR's thermal emission. {A visual 
inspection of radio maps should show whether an \mbox{H\,{\sc ii}} region overlaps with an SNR, so that it could be avoided.} 
We treat such cases with caution, but point out that the existence of the contamination by an \mbox{H\,{\sc ii}} region 
emission does not rule out thermal bremsstrahlung emission from the SNR, it can just mask it.

Another important problem is that our model is sensitive to small changes in the
observations. Deriving accurate integrated flux densities is in fact difficult. It
depends on details of individual observations (e.g.\@ definition of the area that was
integrated or usage of a simplistic Gaussian model; background emission correction; whether
the observations are on the same flux density scale, and whether flux densities are
comparable, i.e.\@ integrated from the same regions). {There is also a problem with total flux densities taken from the 
literature. Some of the authors subtract the contribution from point sources others do not. Some authors even cannot 
subtract them because the resolution of their maps is not good enough. That is in particular a problem at low frequencies. 
Since these point sources are usually extragalactic in origin they could - if not subtracted - steepen the total spectrum 
towards lower frequencies which would make it appear as if the spectrum is concave upwards towards higher frequencies.}

The small number of data points and the dispersion of the flux densities at the same frequencies make for uncertain fitting 
with our model. {More data at radio frequencies, higher than 1 GHz, are necessary in order to make firmer conclusions 
about the significance of thermal emission.}

{Despite these drawbacks, this model can give us rough insight into whether or not
there might be a significant contribution of thermal bremsstrahlung component to the total
volume emissivity at radio frequencies.}

Possible thermal bremsstrahlung radiation at radio frequencies from SNRs was
discussed by Uro{\v s}evi{\' c} \& Pannuti (2005). A similar model, but in the case of spiral
galaxies, was introduced by Duric, Bourneuf \& Gregory (1988).

{There are other important tracers of the thermal emission that could be used to support or dismiss the 
presence of thermal radio emission from SNRs. Linear polarization measurements give us lower limits for the 
non-thermal component of the radio emission. Detection in $\mathrm{H}\alpha$ (see Stupar \& Parker 2011) as well as 
radio recombination lines, implies that a significant amount of thermal electrons must be present. Weak radio 
recombination line emission has been observed toward several supernova remnants and it has remained unclear if this emission 
is in fact associated with SNRs or due to intervening sources such as extended \mbox{H\,{\sc ii}} envelopes along the line 
of sight (Hewitt \& Yusef-Zadeh 2006).
}

\section{Analysis and results}

We {have investigated the radio continuum spectra of the 24 known mixed-morphology SNRs (Vink 2012). We also investigated 
shell and composite SNRs known to expand in high density environments with some of tracers of possible thermal 
emission: curved spectrum, thermal absorption at low frequencies intrinsic to an SNR, detection in radio recombination lines 
and $\mathrm{H}\alpha$, spectral index variations. We used only flux densities with errors $\leq20\%$ from the literature, 
and used only those flux densities that are on the scale of Baars et al.~(1977). We did not analyze the SNRs 
with only four or less data points found.

In this work we discuss the radio continuum properties of 3 SNRs that are the best candidates for testing our hypothesis 
of significant thermal emission.}

\subsection{IC443 (G189.1+3.0) and 3C391 (G31.9+0.0)}

IC443 and 3C391 both expand in a high density environment (and interact with molecular clouds) and exhibit 
thermal absorption at low frequencies which is linked to the SNRs (Brogan 2005, Castelletti et al.~2011).

\subsubsection{IC443}

From our model fit to IC443 (Table 2 and Figure 6), the contribution of thermal emission at 1 GHz 
is between 3 and 57\% (the range of frequencies used here is from 408 MHz to 8 GHz). We used the data from 
Castelletti et al.~(2011) and Gao et al.~(2011). $\chi_{\mathrm{red}}^{2}$ represents the 
$\chi^{2}/\mathrm{dof}$, where dof is the degree of freedom. Adj.\@ $R^{2}$ represents the fit quality\footnote{Scatter 
of residuals relative to the best fit line (adjusted coefficient of determination).}. If we add all the data from Table 2 of 
Castelletti et al.~(2011) with errors less than 20\%, excluding the point at 10 MHz (range of frequencies between 20 MHz 
to 8 GHz), from our model fit, we obtain that there is no significant thermal emission from IC443. {More data, especially between 
10 and 20 GHz are needed to make firm conclusions about the significance of thermal emission.}

\begin{figure}[!t]\centering
  \includegraphics[width=\columnwidth]{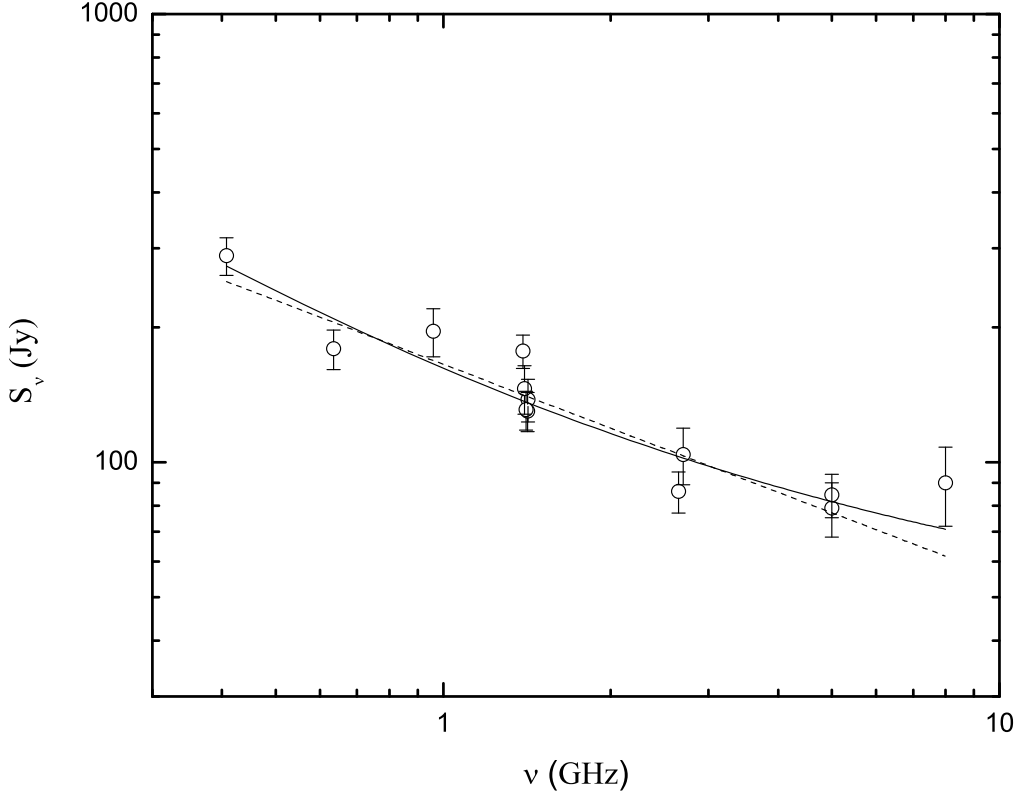}
  \caption{The integrated spectrum of SNR IC443. The full line represents the least
squares fit by the non-thermal plus thermal model, while the dotted line represents the fit
by the purely non-thermal model. Only data on the scale of Baars et al.~(1997) were used.}
  \label{fig:six}
\end{figure}

\begin{table}
\begin{center}
  \caption{Fit parameters for our model and a purely non-thermal model for SNR IC443.} \label{tab:second}
 \begin{tabular}{ccccc}
    \tableline\tableline
    $\alpha$ & $S_{1\ \!\!\mathrm{GHz}}^{\mathrm{NT}}\ \mathrm{(Jy)}$ & $\frac{S_{1\ \!\!\mathrm{GHz}}^{\mathrm{T}}}{S_{1\
\!\!\mathrm{GHz}}^{\mathrm{NT}}}$ &$\chi^{2}/\mathrm{dof}$&Adj.\@ $R^{2}$\\
    \tableline
    $0.82\pm0.35$ & $96.40\pm40.50$ & $0.68\pm0.65$&1.72&0.83\\
    \tableline
    $0.47\pm0.06$ & $165.40\pm7.80$&-&1.74&0.83\\
    \tableline
  \end{tabular}
\end{center}
\end{table}

From the analysis of thermal absorption properties, in the case of IC443 ({eastern rim}), assuming 
$\Omega_{S}=10^{-4}$ sr, $T_{e}=8000-12000$ K and $\nu_{T}=30-50$ MHz, and using equations (8) and (9) we find 
$S_{1\ \!\!\mathrm{GHz}}^{\mathrm{T}}\approx16-68$ Jy. The flux density at 1 GHz, from the power low fit, is around 
160 Jy (Green 2009), which leads to a 10-40\% contribution of thermal emission at 1 GHz. {It must be noted that the 
physical parameters (like $T_{e}$ and $EM$) of the area of thermal absorption, i.e.\@ the eastern half of the SNR 
(Castelletti et al.~2011), were used here.}

{The integrated polarized flux density at 5 GHz for IC443 is $2.6\pm0.3$ Jy or around 3\% for the 
assumed integrated flux density of $84.6\pm9.4$ at 5 GHz (Gao et al.~2011). On the other hand, if we use the results from 
our model fit for non-thermal spectral index (0.82), and synchrotron flux density at 1 GHz (96.40 Jy), we get around 10\% for 
mean linear polarization percentage at 1 GHz. Since the maximum polarization for synchrotron radiation is $\sim$70\%, 
this gives a lower limit for synchrotron emission of around 15\% at 1 GHz, and hence an upper limit for the thermal 
component of approximately 85\%. Of course, one must be aware of the high uncertainties associated with $\alpha$ and 
$S_{1\ \!\!\mathrm{GHz}}^{\mathrm{NT}}$ from our model fit.}

{We also tried to estimate the electron density of the thermal emitting region, from our model, using equations (14) 
and (15), as well as (2). For $d=1.5$ kpc (as in Castelletti et al.~2011), a mean angular SNR dimension of $45'$ (taken from 
Green 2009), and for $T=10^{4}$ K, we get $n_{e}\approx19.5\ f^{-1/2}\ \mathrm{cm^{-3}}$. This is the 
(average) density, smoothed throughout the shell of the volume filling factor $f$. Given the crudeness of our model 
(and the high uncertainty associated with $S_{1\ \!\!\mathrm{GHz}}^{\mathrm{T}}$), as well as high inhomogeneity of 
ISM in the vicinity of this SNR (interaction with molecular cloud), we can not make the firm estimate of $n_{e}$ and 
especially of the swept up mass by the SNR. The crude estimate for the swept up mass is 
$\frac{4\pi}{3}R_{S}^{3}n_{a}m_{H}\approx950\ M_{\odot}$ for a density jump of 4 and $f=0.25$. This is a rather high value. 
It must be noted that this estimate was made by the very crude assumption that the average density of ambient ISM, throughout 
the whole SNR's evolution, is as high as the value we estimated, for the density of thermal emitting region. This estimate of 
the swept up mass is rather the upper limit. As the classical Sedov-Taylor model is not applicable in the case of mixed-morphology 
SNRs, the compression could be much larger than 4 so that the ambient density could be less than $n_{e}/4$. The value of volume 
filling factor could be less than 0.25, too.}

Radio recombination lines have not been detected in the direction of IC443, but only in the direction of adjacent 
\mbox{H\,{\sc ii}} region (Donati-Falchi \& Tofani 1984). IC443 is also detected in H$\alpha$ (Keller et al.~1995). 
{Based on the combination of the new images at 74 and 330 MHz, Castelletti et al.~(2011) investigated spectral
changes with position across IC443, and related them to the spatial characteristics of the radio
continuum emission and of the surrounding ISM. They concluded that thermal absorption at 74 MHz is 
responsible for the localized spectral index flattening observed along the eastern border of IC443. 
Castelletti et al.~(2011) also noted that towards the interior of IC443 the spectrum is consistent with those expected 
from linear DSA, while the flatter spectrum in the southern ridge is a consequence of the strong shock/molecular cloud 
interaction. Electron density roughly estimated for the free-free absorbing gas is $\sim500\ \mathrm{cm^{-3}}$ (Castelletti et al.~2011 
and references therein). Troja, Bocchino, \& Reale (2006) made a rough estimation of the total soft X-ray emitting mass of 
around $30\ M_{\odot}$. They also noted that the material emitting in the X-rays is just the outskirts of large clouds. 
The mean density of the X-ray emitting plasma is around $2.5\ \mathrm{cm^{-3}}$, and the shock velocity corresponding to 
plasma temperatures of 0.3 keV is around $450$ km/s (Troja et al.~2006). The SNR age is around 4 kyr (Troja et al.~2008).}

\subsubsection{3C391}

From our model fit from 1 GHz to 15.5 GHz (Table 3 and Figure 7), we have 10-25\% for the contribution of thermal emission 
at 1 GHz. We used data from Kassim (1989), Brogan et al.~(2005), and Sun et al.~(2011). If we use flux
densities at frequencies higher than 150 MHz (Kassim 1989, Moffett \& Reynolds 1994, Brogan 2005, Sun et al.~2011), from 
our model fit, we obtain that there is no {significant} thermal emission from 3C391. {More data, especially between 
10 and 30 GHz, are needed in order to make firm conclusions about the significance of thermal emission.}

\begin{figure}[!t]\centering
  \includegraphics[width=\columnwidth]{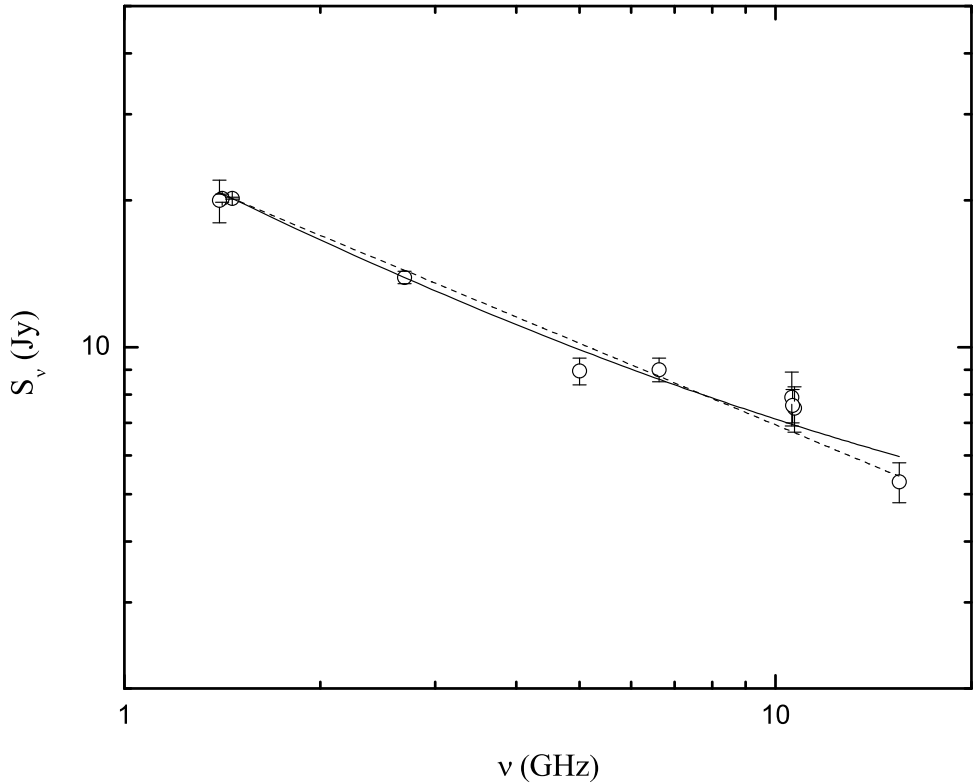}
  \caption{The integrated spectrum of SNR 3C391. The full line represents the least
squares fit by the non-thermal plus thermal model, while the dotted line represents the fit
by the purely non-thermal model. Only data on the scale of Baars et al.~(1997) and frequencies higher than 1 GHz were used.}
  \label{fig:seven}
\end{figure}

\begin{table}
\begin{center}
  \caption{Fit parameters for our model and a purely non-thermal model for SNR 3C391.} \label{tab:second}
 \begin{tabular}{ccccc}
    \tableline\tableline
    $\alpha$ & $S_{1\ \!\!\mathrm{GHz}}^{\mathrm{NT}}\ \mathrm{(Jy)}$ & $\frac{S_{1\ \!\!\mathrm{GHz}}^{\mathrm{T}}}{S_{1\
\!\!\mathrm{GHz}}^{\mathrm{NT}}}$ &$\chi^{2}/\mathrm{dof}$&Adj.\@ $R^{2}$\\
    \tableline
    $0.80\pm0.20$ & $21.18\pm1.41$ & $0.22\pm0.11$&1.34&0.99\\
    \tableline
    $0.56\pm0.02$ & $24.92\pm0.28$&-&1.57&0.99\\
    \tableline
  \end{tabular}
\end{center}
\end{table}

From the analysis of thermal absorption properties, in the case of 3C391, assuming $\Omega_{S}=10^{-6}$ sr, $T_{e}=1000-3000$ K 
and $\nu_{T}=40-150$ MHz (Green 2009, Brogan et al.~2005), using equations (8) and (9), we have $S_{1\ \!\!\mathrm{GHz}}^{\mathrm{T}}\approx0.04-1.7$ Jy. 
The flux density at 1 GHz, from the power low fit, is around 24 Jy (Green 2009), which leads to a 0.15-7\% contribution of thermal emission at 1 GHz.

{The mean polarization fraction of 3C391 is less than 1\% at 5 GHz (Moffett \& Reynolds 1994). This gives the upper limit 
for the thermal component of approximately 98\% at 1 GHz (or lower limit for synchrotron emission of around 2\% at 
1 GHz).}

{For $d=8$ kpc (Chen et al.\@ 2004 and references therein), a mean angular SNR dimension of $\approx6'$ (Green 2009), and 
for $T=10^{4}$ K, we get $n_{e}\approx46\ f^{-1/2}\ \mathrm{cm^{-3}}$ using the value for
$S_{1\ \!\!\mathrm{GHz}}^{\mathrm{T}}$ from our model fit. This is the (average) density, 
smoothed throughout the shell of the volume filling factor $f$. The crude estimate for the swept up mass is 
$\approx806\ M_{\odot}$ for a density jump of 4 and $f=0.25$. As we already pointed out, this value can be interpreted more as 
an upper limit. Also, the classical Sedov-Taylor model is not applicable in the case of mixed-morphology SNRs, so 
the compression could be much larger than 4 and $f$ could be less than 0.25.}

Radio recombination lines have also been observed in the direction of 3C391. Goss et al.~(1979) concluded that the thermal 
component responsible for the low-frequency absorption and radio recombination lines in direction of 3C391 are due to very 
extended, low-density \mbox{H\,{\sc ii}} regions located along the line of sight to 3C391. On the other hand, as we already 
mentioned, Brogan et al.~(2005) found that thermal absorption could be linked to the SNR. {They concluded that the 
free-free absorption originates from the SNR/molecular cloud shock boundaries due to ionized gas created from the 
passage of a J-type shock with a speed of $\sim100$ km/s. Moffet \& Reynolds (1994) did not find convincing evidence for 
spectral index variations above their detection limit of $\Delta\alpha=0.1$. On the other hand, Brogan et al.~(2005) noted 
that the 330/1465 MHz spectral index map is quite uniform, but the 74/330 MHz one shows such spectral behavior that is 
indicative of free-free absorption at 74 MHz. Electron density of the free-free absorbing gas likely lies between $10$ and 
$10^{3}\ \mathrm{cm^{-3}}$, with a lower range between $10$ and $40\ \mathrm{cm^{-3}}$ favored (Brogan et al.~2005).} 
Su \& Chen (2008) pointed out that the mid-IR emission, in the direction of 3C391, is dominated 
by the contribution of the shocked dust grains, which may have been partly destroyed by sputtering. {Chen et al.~(2004) 
suggested the SNR age of around 4 kyr, the mass of the X-ray emitting hot-gas around $114\ f_{x}^{1/2}\ M_{\odot}$, and its 
mean number density of $1.9\ f_{x}^{-1/2}\ \mathrm{cm^{-3}}$, for the most probable SNR distance of 8 kpc, where $f_{x}$ is 
the volume filling factor of the hot-gas. Using the plasma temperature measured from the entire SNR of 0.56 keV, the velocity 
of the blast wave is around 680 km/s (Chen et al.~2004).}

For both, IC443 and 3C391, the errors associated with fit parameters of our model are high. As we already mentioned, we use 
our simple model more for qualitative than for quantitative discussion. On the other hand, in the case of IC443,
our estimate of thermal contribution approximately agrees with that from the analysis of thermal absorption. 
In the case of the 3C391 the conclusions are vague as the results from our model fit and analysis of thermal
absorption do not overlap. {The radio spectral indices of both SNRs, from our model fit, are around 0.8, which is a rather 
high value. Steeper spectra ($\alpha>0.5$) can be expected for much weaker shocks (Bell 1978a,b). On the other hand, one must 
be aware of the parameter uncertainties from our model fit (Tables 2 and 3).} The polarization observations for these SNRs also provide possibility of significant influence of thermal component. {It 
must be mentioned that for the low resolution observations, used to determine the percentage polarization, beam 
depolarization effects should be very important so the results must be treated with caution. Generally, tangled or 
disordered magnetic fields in the emitting region of the radio shell may be responsible for depolarizing the radio 
synchrotron radiation, as well as internal Faraday depolarization (Moffet \& Reynolds 1994). We would like to stress the 
importance of high frequency {($>$1GHz)} observations of SNRs, as well as of the reliable, high resolution, 
observations at low frequencies ($<100$ MHz) as they are necessary in order to make firmer conclusions about the 
existence of "radio thermally active" SNRs.}

\subsection{3C396 (G39.2-0.3)}

The spectrum of 3C396 shows a low-frequency turnover, due to thermal absorption, for which there is no firm evidence 
of direct connection with the SNR. On the other hand, this SNR has obvious "concave up" radio spectrum (Figure 8). 
Su et al.~(2011) investigated the molecular environment of 3C396 and suggested that molecular 
clouds at a distance of 6.2 kpc are in physical contact with this SNR. They also established a scenario for this SNR in which 
the SNR collides with a molecular wall in the west, with a southwestern pillar of molecular gas, and it expands in a 
low density region in the east (but there also encounters a clump of backside molecular gas). In multi-wavelength 
morphologies, both the X-ray and radio emissions are bright in the western half and faint in the eastern half of the SNR, 
which is probably because of the large-scale density gradient (Su et al.~2011). Anderson \& Rudnick (1993) discussed 
the spatial spectral index variations in this SNR. They concluded that 
spectral variations do not coincide with features in total intensity, although they found that the region associated 
with the brightest feature in the SNR (segment of the western annular enhancement, region D in Figure 4 of their work) 
has somewhat flatter spectral index than the SNR mean. The SNR is infra-red bright, and the remnant has not been detected optically 
(Scaife et al.~2007 and references therein). Su et al.~(2010) suggested the SNR age of around 3 kyr, mean blast shock velocity around 
$870\ \mathrm{km\ s^{-1}}$, SNR radius around 7 pc, {and the mass of the X-ray emitting hot-gas around 
$70\ f_{x}^{1/2}\ M_{\odot}$}, for the most probable SNR distance of 6.2 kpc, {where $f_{x}$ is the volume filling factor of 
the hot-gas}. They also estimated the density in the radiative shell of around $400\ \mathrm{cm^{-3}}$, interclump density of around 
$1\ \mathrm{cm^{-3}}$, and density in the clump of around $10^{4}\ \mathrm{cm^{-3}}$. The X-ray bright synchrotron 
pulsar wind nebula, surrounding a yet undetected pulsar, is located in the center of the SNR (Olbert et al.~2003).

\subsubsection{The results of our model fit}

We used the data points from Patnaik et al.~(1990), Scaife et al.~(2007) and Sun et al.~(2011) 
between 80 MHz to 33 GHz. The catalogued flux densities for SNR 3C396 are
contaminated by the nearby steep-spectrum pulsar, PSR 1900+0.5, below 30 MHz (Scaife et al.~2007) or 100 MHz
(Patnaik et al.~1990). Olbert et al.~(2003) emphasized that the pulsar wind nebula (PWN), near the center of the SNR,
contributes $\leq1/25$ of the total radio flux density at 1.4 GHz. The parameters of our model and a purely non-thermal model fit 
are shown in Table 4. In Figure 8 the full line represents a fit to the SNR 3C396 data by the non-thermal 
plus thermal model, while the dotted line represents fit by a purely non-thermal model. We have compared our model with the (simple) 
power law model. From the $\chi^{2}$ statistics we conclude that the non-thermal plus thermal model gives a much better fit 
than a pure non-thermal model for SNR 3C396. From our model, the thermal emission in the case of SNR 3C396, contributes 
$35-47\%$ of the total flux density at 1 GHz. Excluding the points below 100 MHz does not change our results significantly. 
{More data, especially between 10 and 30 GHz, are needed to make firm conclusions about the significance of thermal 
emission.}

\begin{figure}[!t]\centering
  \includegraphics[width=\columnwidth]{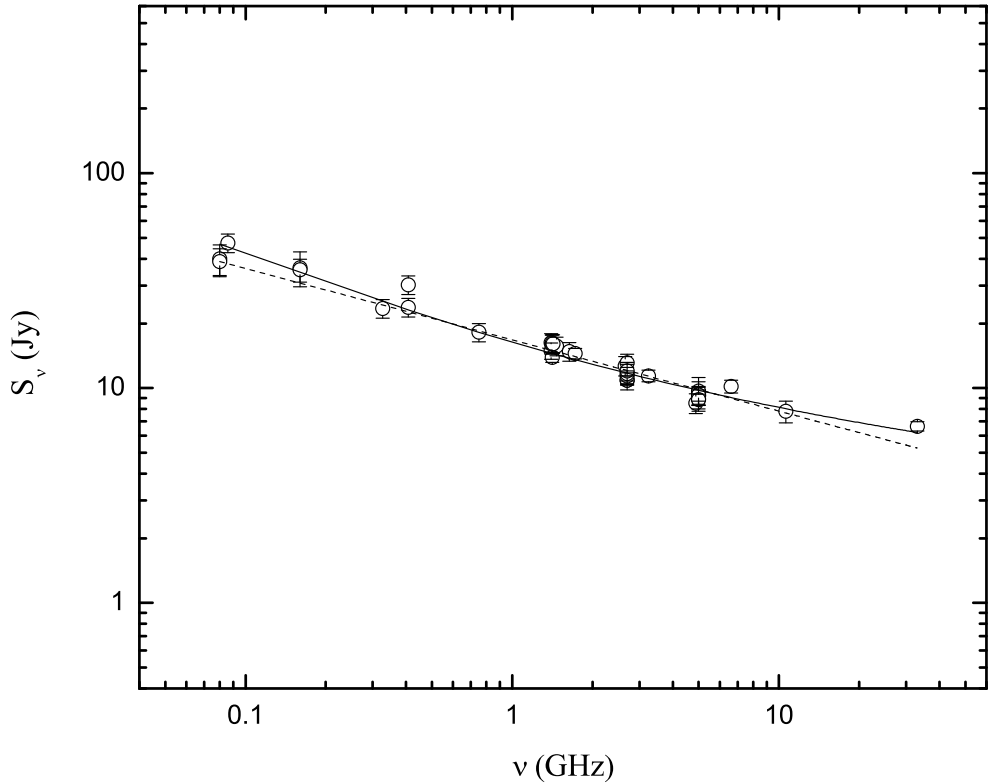}
  \caption{The integrated spectrum of SNR 3C396. The full line represents the least
squares fit by the non-thermal plus thermal model, while the dotted line represents the fit
by the purely non-thermal model.}
  \label{fig:eight}
\end{figure}

\begin{table}
\begin{center}
  \caption{Fit parameters for our model and a purely non-thermal model for SNR 3C396.} \label{tab:second}
 \begin{tabular}{ccccc}
    \tableline\tableline
    $\alpha$ & $S_{1\ \!\!\mathrm{GHz}}^{\mathrm{NT}}\ \mathrm{(Jy)}$ & $\frac{S_{1\ \!\!\mathrm{GHz}}^{\mathrm{T}}}{S_{1\
\!\!\mathrm{GHz}}^{\mathrm{NT}}}$ &$\chi^{2}/\mathrm{dof}$&Adj.\@ $R^{2}$\\
    \tableline
    $0.55\pm0.05$ & $9.56\pm1.14$ & $0.71\pm0.18$&1.01&0.94\\
    \tableline
    $0.33\pm0.02$ & $16.78\pm0.38$&-&1.58&0.90\\
    \tableline
  \end{tabular}
\end{center}
\end{table}

\subsubsection{Discussion on a low-frequency turnover}

The low frequency spectrum of SNR 3C396 (see Fig.~10 in Patnaik et al.~1990) shows a turnover below 80-100 MHz 
(Anderson \& Rudnick 1993). For spectral turnover due to thermal absorption around 40-80 MHz, the electron temperature would be order of
$T_{e}\approx(1-6)\times10^{4}$ K (using equation 8) for our model. In that case, the emission measure is
$EM\approx10^{4-5}\ \mathrm{cm^{-6}\ pc}$. On the other hand, our results are not consistent
with the work of Anantharamaiah (1985), who gave an upper limit on the emission measure of 280 $\mathrm{cm^{-6}\ pc}$ and
electron temperature of 5000 K for the gas, which leads to $\nu_{T}\approx20$ MHz and thermal contribution of 0.8\% at 1 GHz. 
If we keep $T_{e}=5000$ K but put $\nu_{T}=40-80$ MHz we have $EM\approx10^{3}\ \mathrm{cm^{-6}\ pc}$ and
not significantly higher thermal contribution of $\approx$0.9\% at 1 GHz.
Anantharamaiah (1985) also noted that the upper limits implied by the pulsar dispersion measure (Table 4 
in the mentioned work) are not rigorous due to the assumptions made. Hewitt et al.~(2009) pointed out, from the {\it Spitzer}
IRS observation analysis, that along the western shell, the inner
[\mbox{Fe\,{\sc ii}}]-line filament and the outer H$_{2}$-line one are found to be spatially separated. Their results are in
a good agreement with the near-IR observations of Lee et al.~(2009). Hewitt et al.~(2009) used
observations of strong ionic lines, and found $n_{e}$=270 $\mathrm{cm}^{-3}$, $T_{e}=2.3\times10^{4}$ K from the
[\mbox{Fe\,{\sc ii}}] 7.9/5.35 $\mathrm{\mu m}$ and 17.9/26 $\mathrm{\mu m}$ ratios. On the other hand, Lee et al.~(2009),
using the Wide-Field Infrared Camera aboard Palomar 5 m Hale telescope, found an upper limit for electron number density of
the [\mbox{Fe\,{\sc ii}}] 1.64 $\mathrm{\mu m}$ emission gas of $<2000\ \mathrm{cm}^{-3}$ for assumed temperature of 5000 K.
The increased abundance of gas phase iron by shock-induced sputtering of the dust grains and/or the creation of an extensive
partially ionized zone by shock heating can substantially enhance the [\mbox{Fe\,{\sc ii}}] emission and also implies that
most of the dust has been destroyed (Lee et al.~2009, and references therein). Given the very complex nature of this remnant
(high density gradients, interaction with molecular clouds), the shape of the integrated radio spectrum from 10-100 MHz, as
well as its multi-wavelength properties, the presence of thermal emission for 3C396 is uncertain. We point out
that spatially resolved low frequency radio observations are needed to make firmer conclusions. {We propose, for a 
future work, the observations of SNR 3C396, with one, VLA\footnote{The Very Large Array of the National Radio Astronomy 
Observatory is a facility of the National Science Foundation operated under cooperative agreement by Associated Universities, 
Inc.}, telescope over a wide frequency range. That would also include obtaining the high resolution radio images at 74 and 
330 MHz for sensitive, spatially resolved, spectral analysis of the radio emission at long wavelengths.}

The bright 24 $\mathrm{\mu m}$ mid-IR emission, in the direction of 3C396, may chiefly come from dust grains and even 
from the ionic and molecular species in the shocked gas produced when the western molecular wall was hit by the SNR 
blast wave (Su et al.~2011, and references therein). Scaife et al.~(2007) have assessed the possibility of spinning dust 
emission at 33 GHz towards the SNR 3C396. They also concluded that further measurements are required in the range 10-20 GHz 
in order to confirm that. These measurements will also help determine if there is significant thermal emission from 3C396.

\subsubsection{Linear polarization measurements}

From a linear polarization analysis at 5 GHz, Patnaik et al.~(1990) state that the polarization percentages towards the 
eastern edge of the SNR reach about 50\%. On the other hand, on the SW and NW shell structure the polarization is typically 
6-7\%. The central polarized island is about 10\% polarized, while the rest of the interior has a percentage polarization of 
$\leq3$\% (Patnaik et al.~1990). Sun et al.~(2011) found an integrated polarized flux density at 5 GHz $240\pm15$ mJy. The
average percentage polarization is 3\% if we assume the integrated flux density at 5 GHz of $8.84\pm0.53$ Jy 
from their work. {On the other hand, if we use the results from our model for non-thermal spectral index (0.55) and 
synchrotron flux density at 1 GHz (9.56 Jy), we get around 6\% for mean linear polarization percentage at 1 GHz. This gives 
an upper limit for the thermal component of approximately 91\% at 1 GHz (or lower limit for synchrotron emission of around 
9\% at 1 GHz). Of course, as we already noted, one must be aware of the uncertainties associated with $\alpha$ and 
$S_{1\ \!\!\mathrm{GHz}}^{\mathrm{NT}}$, from our model fit.}

\subsubsection{The density and swept up mass estimation}

F{or $d=6.2$ kpc (as in Su et al.~2011), a mean angular SNR dimension of $\approx7'$ (Green 2009), and 
for $T=10^{4}$ K we get $n_{e}\approx50\ f^{-1/2}\ \mathrm{cm^{-3}}$ using the value for
$S_{1\ \!\!\mathrm{GHz}}^{\mathrm{T}}$ from our model fit. This is the (average) density, 
smoothed throughout the shell of the volume filling factor $f$. Given the crudeness of our model, as well as 
high inhomogeneity of ISM in the vicinity of this SNR (interaction with molecular cloud), we can not make a firm 
estimate of $n_{e}$ and especially of the swept up mass. The crude estimate of swept up mass is 
$\approx650\ M_{\odot}$ for a density jump of 4 and $f=0.25$. This value can be interpreted more as 
an upper limit. Such an estimate was made by the very crude assumption that the average density of ambient ISM, 
throughout the whole SNR's evolution, is as high as the value we estimated, for the density of thermal emitting region.}

\section{Conclusions}

In this work, we investigated the possibility of significant production of thermal
bremsstra\-hlung radiation at radio continuum frequencies intrinsic to some Galactic SNRs.

{(1) There are several possible tracers of significant radio thermal bremsstra\-hlung radiation from SNRs,
such as: a "concave up" radio spectrum (flattening at higher frequencies), thermal absorption at lower frequencies intrinsic
to the SNR, and flat spectral indices (radio spectral index variations) associated with high density regions which do not show 
high linear polarization.

(2) The main targets for investigation of the presence of significant radio thermal emission are SNRs expanding in 
high density as well as in inhomogeneous environments such as those interacting with molecular clouds. The evolution as well 
as the radiation from such SNRs is more complex than usually considered in models such as for Sedov-Taylor evolution.

(3) In this work we discussed the radio continuum properties of 3 SNRs that are good candidates for testing our 
hypothesis on significant thermal emission. In the case of the SNRs IC443 and 3C391, thermal absorption, linked to the
SNRs was previously detected. For the IC443, the contribution of the thermal emission (3-57\% at 1 GHz) approximately 
agrees with the estimate obtained from low-frequency thermal absorption (10-40\% at 1 GHz). For SNR 3C391 the
results of our model fit (10-25\% at 1 GHz) have higher values than those from the analysis of thermal absorption
(0.15-7\% at 1 GHz). In the case of the SNR 3C396 we suggest that thermal absorption could be linked to the SNR and propose
that the thermal emission ($<$47\% at 1 GHz from our model fit) could be significant enough to shape the radio continuum
spectrum at high frequencies. The polarization observations for these SNRs also allow
the presence of a significant thermal component.

(4) The small number of data points (with acceptable associated errors) and the
dispersion of flux densities at the same frequencies prevent us from firm quantitative analysis. {Reliable 
observations at low frequencies ($<100$ MHz), as well as more data at radio frequencies higher than 1 GHz are 
necessary in order to make stronger conclusions about the existence of "radio thermally active" SNRs.}

{(5) We propose, for a future work, a systematic survey of candidate "radio thermally active" SNRs, especially 
SNR 3C396, with one telescope (the eVLA) over a wide frequency range. That would also include obtaining high resolution 
radio images at 74 and 330 MHz for sensitive, spatially resolved, spectral analysis of the radio emission at long 
wavelengths.
}

(6) The analysis of possible thermal bremsstrahlung radio emission inherent to some Galactic SNRs could be very important
tool for the estimation of density in which Galactic or extragalactic SNRs are embedded.
}

\section*{ACKNOWLEDGEMENT}

We wish to thank the anonymous referee for useful suggestions which substantially improved this paper. 
This work is part of the projects 176005 
"Emission nebulae: structure and evolution" supported by the Ministry of Education and Science of Serbia. D.A.L. 
acknowledges support from the Natural Sciences and Engineering Research Council of Canada.


\begin{thebibliography}

\bibitem[Allen, Houck \& Sturner(2008)]{2008ApJ...683..773} Allen, G.~E.,
Houck, J.~C., \& Sturner, S.~J.\ 2008, \apj, 683, 773

\bibitem[Altenhoff et al.(1970)]{1970AAS...1..319A} Altenhoff, W.~J., Downes,
D., Goad, L., Maxwell, A., \& Rinehart, R.\ 1970, \aaps, 1, 319A

\bibitem[Altenhoff et al.(1960)]{1960VSB...59} Altenhoff, W.~J., Mezger, P.~G., Wendker, H., Westerhout, G.\ 1960, Ver$\mathrm{\ddot{o}}$ff. Sternw$\mathrm{\ddot{a}}$rte Bonn, 59, 48

\bibitem[Anantharamaiah (1985)]{1985AA...203} Anantharamaiah, K.~R.\ 1985, \aap, 6, 203

\bibitem[Anderson \& Rudnick (1993)]{1993ApJ...514} Anderson, M.~C., \& Rudnick, L.\ 1993, \apj, 408, 514

\bibitem[Baars et al.(1977)]{1977AA...61..99B} Baars, J.~W.~M., Genzel, R.,
Paulinu-Toth, I.~I.~K., \& Witzel, A.\ 1977, \aap, 61, 99B

\bibitem[Bandiera, \& Petruk(2004)]{2004AA...419..419} Bandiera, R., \&
Petruk, O.\ 2004, \aap, 419, 419

\bibitem[Bell(1978)a]{1978MNRAS...182..443} Bell, A.~R.\ 1978a, \mnras, 182, 147

\bibitem[Bell(1978)b]{1978MNRAS...182..443} Bell, A.~R.\ 1978b, \mnras, 182, 443

\bibitem[Bell, Schure \& Reville(2011)]{2011MNRAS..418...1208} Bell, A.~R., Schure, K.~M., \& Reville, B.\ 2011, \mnras, 418, 1208

\bibitem[Berezhko \& V\"{o}lk(2004)]{2004..BV} Berezhko, E.~G., \& V\"{o}lk, H.~J.\ 2004, \aap, 427, 525

\bibitem[Brogan et al.(2005)]{2005..Broganetal} Brogan, C.~L., Lazio, T.~J., Kassim, N.~E., \& Dyer, K~K.\ 2005, \aj, 130, 148

\bibitem[Castelletti et al.(2011)]{2011..Cas} Castelletti, G., Dubner, Clarke, G.~T., \& Kassim, N.~E.\ 2011, \aap, 534, 21

\bibitem[Chen et al.(2004)]{2004..CHEN} Chen, Y., Su, Y., Slane, P.~O., \& Wang, Q.~D.\ 2004, \apj, 616, 885

\bibitem[Chomiuk \& Wilcots(2009)]{2000..CW} Chomiuk, L., \& Wilcots, E.~M.\ 2009, \apj, 703, 370

\bibitem[Cooray \& Furlanetto(2004)]{2004ApJL...606..L5} Cooray, A., \&
Furlanetto, S.~R.\ 2004, \apjl, 606, L5

\bibitem[]{} Donati-Falchi, A., \& Tofani, G.\ 1984, \aap, 140, 395 

\bibitem[Drury et al.(2009)]{2009..Drury} Drury, L.~O'C., Aharonian, F.~A., Malyshev, D., \& Gabici S.\ 2009, \aap, 496, 1D

\bibitem[Duric, Bourneuf \& Gregory(1988)]{1988AJ....96..81} Duric, N.,
Bourneuf, E., \& Gregory P. C.\ 1988, \aj, 96, 81

\bibitem[Gao et al.(2011)]{2011..Gao} Gao, X.~Y., Han, J.~L., Reich, W., Reich, P., Sun, X.~H., Xiao, L.\ 2011, \aap, 529, 159

\bibitem[Gayet(1970)]{1970AA..9..312G} Gayet, R.\ 1970, \aap, 9, 312

\bibitem[]{} Goss, W.~M., Skellern, D.~J., Watkinson, A., Shaver, P.~A.\ 1979, \aap, 78, 75

\bibitem[Green(2007)]{2007BASI..35...77} Green, D.~A.\ 2007, BASI, 35, 77

\bibitem[Green(2009)]{2009...Greens..catalogue} Green, D.~A.\ 2009, A
catalogue of Galactic Supernova Remnants (2009 March version), Astrophysics
Group, Cavendish Laboratory, Cambridge, United Kingdom (available at
http://www.mrao.cam.ac.uk/surveys/snrs/)

\bibitem[Hewitt et al.(2009)]{2009apj..694...1266} Hewitt, J.~W., Rho, J, Andersen, M., \&, and Reach, W.~T.\ 2009, \apj, 694, 1266

\bibitem[]{} Hewitt, J.~W., \& Yusef-Zadeh, F.\ 2006, AAS, 208, 4705H

\bibitem[Kassim(1989)]{1989..Kas} Kassim, N., E.\ 1989, \apj, 71, 799

\bibitem[]{} Keller, L.~D., Jaffe, D.~T., Pak, S., Luhmal, P.~L., \& Claver, C.~F.\ 1995, RevMexAA, 3, 251

\bibitem[Lahy(2006)]{2006..Leahy} Leahy, D.~A.\ 2006, \apj, 647, 1125

\bibitem[Leahy \& Roger(1998)]{1998..LR} Leahy, D.~A., \& Roger, R.~S.\ 1998, \apj, 505, 784

\bibitem[]{}Leahy, D.~A., \& Tian, W.~W.,\ 2006, \aap, 451, 251

\bibitem[]{} Leahy, D.~A., Xizehn, Z., Xinji, W., \& Jiale, L.\ 1998, \aap, 339, 601

\bibitem[]{} Lee, Ho-Gyu, Moon, Dae-Sik, Koo1, Bon-Chul, Lee, Jae-Joon, \&  Matthews, K.\ 2009, \apj, 691, 1042

\bibitem[]{} Moffett, D.~A., Reynolds, S.~P.\ 1994, \apj, 425, 668

\bibitem[McKee \& Ostriker(1977)]{1977...218..148} McKee, C.~F., \& Ostriker,
J.~P.\ 1977, \apj, 218, 148

\bibitem[Olbert et al.(2003)]{2003ApJL...592..L45} Olbert, C., Keohane, J.~W.,
Arnaud, K.~A., Dyer, K.~K., Reynolds, S.~P., \& Safi-Harb, S.\ 2003, \apjl,
592, L45

\bibitem[Oni\'c \& Uro{\v s}evi{\'c}(2008)]{2008SAJ...177..59} Oni\'c, D., \&
Uro{\v s}evi{\' c}, D.\ 2008, Serb. Astron. J., 177, 59

\bibitem[Ostrowski(1999)]{1999..OS} Ostrowski, M.\ 1999, \aap, 345, 256

\bibitem[Patnaik et al.(1990)]{1990AA...232..467} Patnaik, A.~R., Hunt, G.~C.,
Salter, C.~J., Shaver, P.~A., \& Velusamy, T.\ 1990, \aap, 232, 467

\bibitem[Petruk(2005)]{2005...9..364P} Petruk, O.\ 2005, JPhSt, 9, 364P

\bibitem[Pynzar' \& Shishiv(2007)]{2007..PS} Pynzar' A.~V., \& Shishiv, V.~I.\ 2007, Astron.~Rep., 51, 1

\bibitem[Reynolds(2008)]{2008ARAA..46...89} Reynolds, S.~P.\ 2008, \araa, 46, 89

\bibitem[Reynolds(2011)]{2011ApSS..336...257R} Reynolds, S.~P.\ 2011, \apss, 336, 257

\bibitem[Reynolds \& Ellison(1992)]{1992ApJ...399..L75} Reynolds, S.~P., \&
Ellison, D.~C.\ 1992, \apj, 399, L75

\bibitem[Rho \& Petre(1998)]{2008..RP} Rho, J., \& Petre, R.\ 1998, \apj, 503, L167

\bibitem[Scaife et al.(2007)]{2007MNRASL...377..L69} Scaife, A., Green, D.~A.,
Battye, R.~A., Davies, R.~D., Davis, R.~J., Dickinson, C., Franzen, T.,
G$\mathrm{\acute{e}}$nova-Santos, R., Grainge, K., Hafez, Y.~A., and 11
coauthors\ 2007, \mnras, 377, L69

\bibitem[Schlickeiser \& F\"{u}rst(1989)]{1989AA...219..192} Schlickeiser,
R., \& F\"{u}rst, E.\ 1989, \aap, 219, 192

\bibitem[Shi et al.(2008)]{2008AA...487..601} Shi, W.~B., Han, J.~L., Gao,
X.~Y., Sun, X.~H, Xiao, L., Reich, P., \& Reich, W.\ 2008, \aap, 487, 601

\bibitem[Stupar \& Parker 2011]{2011..MNRAS..414..2282S} Stupar, M., Parker, Q.~A.\ 2011, \mnras, 414, 2282S

\bibitem[Su et al.(2011)]{2011..SU} Su, Y., Chen, Y., Koo, Bon-Chul, Zhou, X., Lu, Deng-Rong, Jeong, Il-Gyo, \& DeLaney, T.\ 2011, \apj, 727, 43

\bibitem[]{} Su, Y., \& Chen, Y.\ 2008, Adv.~Space Res., 41, 401

\bibitem[]{} Sun, X.~H., Reich, P., Reich, W., Xiao, L., Gao, X.~Y., \& Han, J.~L.\ 2011, \aap, 536A, 83S

\bibitem[Tian \& Leahy(2005)]{TA..2005} Tian, W.~W., \& Leahy, D.~A.\ 2005, \aap, 436, 187

\bibitem[Tilley, Balsara \& Howk(2006)]{2006MNRAS...371..1106} Tilley, D.~A.,
Balsara, D.~S., \& Howk, J.~C.\ 2006, \mnras, 371, 1106

\bibitem[Troja et al.(2008)]{2008..Troj} Troja, E., Bocchino, F., Miceli, M., \& Reale, F.\ 2008, \aap, 485, 777

\bibitem[Troja et al.(2006)]{2006..Troj} Troja, E., Bocchino, F., \& Reale, F.\ 2006, \apj, 649, 258

\bibitem[Truelove \& McKee(1999)]{1999ApJS...120..299} Truelove, J.~K., \&
McKee, C.~F.\ 1999, \apjs, 120, 299

\bibitem[Uchiyama et al.(2010)]{2010..apjl..723..L122} Uchiyama, Y., Roger D. Blandford, R.~D., Funk, S., Tajima, H., \& Tanaka, T.\ 2010, \apjl, 723, L122

\bibitem[Uro{\v s}evi{\' c} \& Pannuti(2005)]{2005ASPPH...23..577} Uro{\v s}evi{\' c}, D., \& Pannuti, T.~G.\ 2005, Astropart. Phys., 23, 577

\bibitem[Uro{\v s}evi{\' c}, Pannuti \& Leahy(2007)]{2007ApJL...655..L41} Uro{\v s}evi{\' c}, D., Pannuti, T.~G., \& Leahy, D.\ 2007, \apjl, 655, L41

\bibitem[Vink et al.(2010)]{2010..Vinketal.} Vink, J., Tamazaki, R, Helder, E.~A, \& Schure, K.~M.\ 2010, \apj, 722, 1727

\bibitem[Vink (2012)]{2012..Vink} Vink, J.\ 2012, Astron Astrophys Rev, 20, 49

\bibitem[Yamauchi et al.(1999)]{1999PASJ...51..13} Yamauchi, S., Koyama, K.,
Tomida, H., Yokogawa, J., \& Tamura, K.\ 1999, PASJ, 51, 13

\bibitem[Yamaguchi et al.(2012)]{2012...49..451} Yamaguchi, H., Ozawa, M.,
Ohnishi, T.\ 2012, Adv.~Space Res., 49, 451

\end{thebibliography}
\end{document}